\documentclass[aps,prd,preprint,showpacs]{revtex4-1}
\usepackage{graphicx}
\usepackage{amsmath}
\usepackage{caption}
\usepackage{slashed}

\begin{document}


\title{Vector meson-Baryon dynamics and generation of resonances.}


\author{K. P. Khemchandani$^1$\footnote{kanchan@rcnp.osaka-u.ac.jp}}
\author{H. Kaneko$^1$\footnote{kaneko@rcnp.osaka-u.ac.jp}}
\author{ H. Nagahiro$^2$\footnote{nagahiro@rcnp.osaka-u.ac.jp} }
\author{ A. Hosaka$^1$\footnote{hosaka@rcnp.osaka-u.ac.jp}}

 \affiliation{$^1$Research Center for Nuclear Physics (RCNP), Mihogaoka 10-1, Ibaraki 567-0047, Japan. \\
$^2$Department of Physics, Nara Women's University,  Nara 630-8506, Japan.}


\date{\today}

\begin{abstract}
The purpose of this work is to study vector meson-octet baryon interactions with the aim to find dynamical generation of 
resonances in such systems. For this, we consider s-, t-, u-channel diagrams along with a contact interaction originating from the hidden local symmetry Lagrangian. We find the contribution from all these sources, except the s-channel, to be important. The amplitudes obtained by solving coupled channel  Bethe Salpeter equations for systems with total strangeness zero, show generation of one isospin 3/2, spin 1/2 resonance and three isospin 1/2 resonances: two with spin 3/2 and one with spin 1/2. We identify these resonances with $\Delta (1900)~S_{31}$, $N^* (2080)~D_{13}$, $N^* (1700)~D_{13}$, and  $N^*(2090)~S_{11}$, respectively. 
\end{abstract}

\pacs{  }

\maketitle

\section{Introduction}
Recent interests in hadron physics have been largely motivated by experimental observations of new states in the resonance region which are not easily explained by the conventional constituent quark model.  The strong interactions among the ground state mesons and baryons not only affect their properties but also in some cases generate resonances  dynamically  (examples of some of the recent related works are Refs.~\cite{dgr1,dgr2,dgr3,dgr4,dgr5,dgr6,dgr7}).  
Therefore, it is of great importance to investigate these dynamical aspects based on reliable hadron-hadron interactions. 

In a quark picture, an energy of several hundred MeV which is a typical scale of one quanta of orbital excitation is sufficient to create a $\bar{q}q$ pair, making multiquark components in a hadron.  
If they further develop color singlet clusters of ground state hadrons near their threshold, they may form a loosely bound or resonant state provided that sufficiently strong attraction is available.  
This is what we expect microscopically for the dynamical generation of resonances.  
A spin zero configuration of $\bar{q}q$ forms a $J^P = 0^-$ pseudoscalar meson, and is the basic building block of, for instance, $\Lambda(1405)$ 
\cite{kaiser,ramos,jidocollab}.  
Similarly, the $J^P = 1^-$ configuration giving a vector meson could also be an element of certain baryon resonances as indicated in Refs. \cite{juelich,lutz,pedro,michael,sourav,eulogio}.  
However, while the pseudoscalar meson baryon interaction is well dictated by the low energy theorems of spontaneously broken chiral symmetry, the interaction of vector mesons and baryons are not fully studied.  
This is one of the issues that we would like to discuss in this paper.  

It is known that the theory of the hidden local symmetry (HLS)  \cite{bando} can accommodate  vector mesons consistently with the chiral symmetry. 
In fact, the HLS model has been shown to share many important aspects of low energy dynamics.    
Furthermore, a recent holographic approach to QCD has derived the extended HLS model where infinite series of the vector mesons emerge as a consequence of the dynamics in the extra fifth dimension \cite{holographicqcd1,holographicqcd2}. This HLS model forms the basis of our study

The vector meson-octet baryon interaction has been studied within the HLS, by assuming a  vector meson exchange in the t-channel \cite{eulogio} (Fig.~\ref{diagram}a)
 as the lowest order amplitude and several baryon resonances have been found as a result of solving the Bethe Salpeter equation
 in the coupled channel formalism. However, in Ref. \cite{eulogio}
all the states are found to be spin 1/2-3/2 degenerate since the leading order interaction obtained from the t-channel exchange is spin 
independent. This latter finding is different from what one would expect from the interaction of two particles of similar mass and non-zero spin,  just as for the nuclear force.
In addition to this, the low energy theorems cannot be applied to the vector meson-baryon systems since the masses of the vector mesons are
comparable to those of the baryons. Hence one cannot rely alone on the t-channel diagrams as in the case of pseudoscalar meson- baryon
interaction. There are many other diagrams which could also make important contributions to the interaction of the vector mesons and baryons
and there is no apriori reason to neglect such diagrams, for instance, s- and u-channel (baryon) exchange (Fig.~\ref{diagram}c, d) or a contact interaction (Fig.~\ref{diagram}b).

When naively applying the HLS model to the vector-meson baryon interactions, as we will show in this work, one naturally finds  interactions corresponding to all the diagrams of Fig.~1a-d.  
Moreover,  in the diagrams corresponding to the s-, u-, channels and contact interaction, we find a spin-dependence, which seems also a natural consequence for particles with finite spin.  
It is therefore of great importance to see the role of the diagrams other than the t-channel for the dynamics of the system of a vector-meson and a baryon.  
As we will show in detail, such interactions modify substantially the results which are obtained by employing the spin-independent t-channel interaction.  

In the next section, we will briefly summarize the basic Lagrangians obtained within the HLS approach, in the SU(2) limit first since we find it very instructive to look at the different features and structure of the interactions obtained from different diagrams. We will then discuss the generalization of the same to SU(3).  In the subsequent section we will show the results obtained  on the real energy axis and in the complex plane for the vector meson-baryon systems with total strangeness zero. Finally, we will present a summary of this article.

\section{Formalism}\label{two}
\begin{figure}[h]
\begin{center}
\includegraphics[width=13cm]{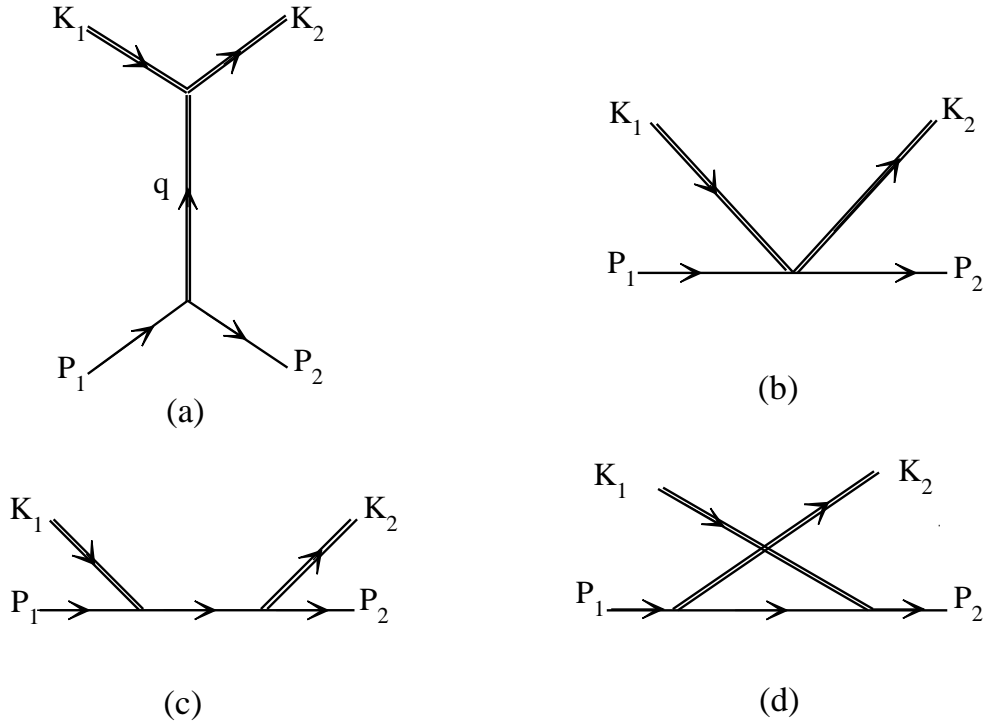}
\caption{ Diagrammatical representation of the vector meson-baryon interaction via a  (a) t-channel exchange, (b) contact term, (c) s-channel and (d) u-channel exchange. The double lines in these diagrams represent the vector mesons.  }
\label{diagram}
\end{center}
\end{figure}

As already mentioned in the previous section, unlike the  pseudoscalar meson-baryon systems, in case of the vector meson-baryon interaction there is not much guidance available from the low energy theorems.  Further, the situation is slightly more complicated since both the particles possess non-zero spin. Thus we make an assumption that the minimal coupling between vector mesons and baryons, based on the HLS model,   occurs via  a
vector meson exchange in the t-channel, a contact interaction and  an octet baryon exchange in the s- and u-channel (shown in Figs.~\ref{diagram} a-d, respectively).  

\subsection{Vector meson-baryon interaction within SU(2)}
Before discussing the interactions in SU(3), we would first 
like to show the structure of the contributions obtained from different diagrams shown in Fig.~\ref{diagram},
in the SU(2) limit which will serve as a guidance in the discussion of  generalization of these interactions to SU(3)  and
in the interpretation of the results finally obtained by solving the scattering equations. 

Our basic assumption is that the nucleon field transforms as  $N \rightarrow h(x) N$ when the hidden local symmetry  is applied to the system, where $h(x)$ is an element of the HLS. Hence the  corresponding gauge invariant HLS Lagrangian is given by
\begin{equation}
\mathcal{L} =  \bar{N} \left( i \slashed \partial - g F_1 \gamma_\mu \rho^\mu \right) N,
\end{equation}
where $F_1(q^2)$ is a form factor normalized as $F(0) = 1$.
However, it is well known that to reproduce the anomalous magnetic moments of the baryons, it is  imperative to consider the rho-nucleon tensor
coupling. This phenomenological finding leads to a more complete $\rho N$ HLS Lagrangian  
\begin{equation}
\mathcal{L}_{\rho N } = - g \bar{N} \left\{ F_1 \gamma_\mu \rho^\mu + \frac{F_2}{4M} \sigma_{\mu\nu} \rho^{\mu\nu} \right\} N, \label{rhoNL}
\end{equation}
where  $F_2 = \kappa_\rho$ at zero momentum transfer, and
\begin{equation}
\rho^{\mu\nu} = \partial^{\mu} \rho^\nu - \partial^{\nu} \rho^\mu + ig \left[\rho^\mu, \rho^\nu \right]. \label{rhotensor}
\end{equation}
It should be stressed here that the commutator term of Eq.~(\ref{rhotensor}) is essential for the gauge invariance of  the tensor term of the $\rho N$ Lagrangian given by Eq.~(\ref{rhoNL}).

We use the Dirac representation of the gamma matrices and in our normalization scheme,
\begin{equation}
\rho^\mu = \frac{\vec{\tau}}{2} \cdot \vec{\rho}.
\end{equation}
Using the above ingredients together with the Lagrangian for three-rho vertices  which comes from the kinetic term of 
the $\rho$ meson 
\begin{equation}
\mathcal{L}_{3\rho} \in -\frac{1}{2} \langle \rho^{\mu\nu} \rho_{\mu\nu} \rangle,
\end{equation}
where $\langle ... \rangle$ denotes a trace in the isospin space, we obtain the  leading order contribution to the vector meson-baryon $T$-matrix, in the SU(2) limit, from a vector exchange in the t-channel
as
\begin{eqnarray}
V^t_{\rho N} = &-\dfrac{1}{2 f_\pi^2} \left( K_1^0 + K_2^0 \right) \vec{\epsilon_1} \cdot \vec{\epsilon_2} & \,\,\,\, {\rm for  \,\,{\it I} = 1/2} \label{t1}\\
= &\dfrac{1}{4 f_\pi^2} \left( K_1^0 + K_2^0 \right) \vec{\epsilon_1} \cdot \vec{\epsilon_2} & \,\,\,\, {\rm  for \,\,  {\it I} =  3/2}. \label{t2}
\end{eqnarray}
In what follows, we shall refer to the leading order contributions to the $T$-matrices simply as interactions (kernels for the Bethe Salpeter equations). The subscripts $1$ and $2$ in the above equations and, in general, in the present discussions refer to the hadron in the initial and final state, respectively (as shown in Figs.~\ref{diagram}a-d). Further, $\vec{\epsilon_i}$ in Eqs.~(\ref{t1}) and (\ref{t2}) denote the polarization vectors of $\rho$, and the energies,  $K_i^0$,  of the
mesons are calculated as
\begin{equation}
K_i^0 = \frac{s + m_i^2 - M_i^2} {2\sqrt{s}},
\end{equation}
where $\sqrt{s}$, $m$ and $M$, here and throughout this article, represent the total energy and masses of the meson and the
baryon, respectively.  
For this t-channel case, we find that the contribution from the  Pauli  term of the Lagrangian for the vector meson-baryon system (Eq.  (\ref{rhoNL})), which
is related to the anomalous magnetic moment, is negligible (as also found in \cite{eulogio}) and hence it can be approximated to 
\begin{equation}
\mathcal{L}_{\rho NN} \simeq - g \bar{N} \gamma_\mu \rho^\mu N.
\end{equation}
However, this anomalous magnetic part plays a very important role in giving another interaction term, a contact term (C.T) when the 
last term of the  tensor field (Eq.~\ref{rhotensor}), which is related to two meson fields, is used in Eq.~(\ref{rhoNL})
\begin{equation}
\mathcal{L}^{C.T}_{\rho N\rho N} = - i g^2 \bar{N}  \dfrac{\kappa_\rho}{2M} \rho_\mu \rho_\nu \sigma^{\mu\nu} N.
\end{equation}
We find that the structure of the $T$-matrix obtained from  this contact interaction, at the lowest order, in the non-relativistic approximation, is
\begin{equation}
V^{C.T}_{\rho N} =   \dfrac{g^2 \kappa_\rho}{M} \vec{t_\rho} \cdot \vec{t_N} \vec{s_\rho} \cdot \vec{s_N},
\end{equation}
which, in contrast to the structure of the interaction proceeding through a t-channel vector meson exchange (summarized from Eqs. \ref{t1} and \ref{t2}) 
\begin{equation}
V^t_{\rho N} =  \dfrac{1}{2 f_\pi^2} \left( K_1^0 + K_2^0 \right)  \vec{t_\rho} \cdot \vec{t_N},
\end{equation}
possesses a spin-spin interaction (see also Eq.~(\ref{vcontact}) and the discussion following it). Further using the KSRF relation  \cite{ksrf1,ksrf2}
\begin{equation}
g=  \frac{m}{\sqrt{2} f_\pi},
\end{equation}
we find that the contact term of the vector meson-baryon HLS Lagrangian is similar, in order of magnitude, to the t-channel vector exchange
contribution. 

Furthermore, using Eq.~(\ref{rhoNL}) we obtain the interactions corresponding to a nucleon exchange in the  s- and u-channel  diagrams, which in the non-relativistic approximation are given as
\begin{eqnarray}
V^{s}_{\rho N} &=& \frac{g^2}{4} \left[ 1 - \dfrac{\kappa_\rho m}{2M}\right]^2 \dfrac{1}{m + 2M}\left(1 - 2 \vec{t_\rho} \cdot \vec{t_N} \right) \left(1 - 2 \vec{s_\rho} \cdot \vec{s_N} \right), \\
V^{u}_{\rho N} &=&  \frac{g^2}{4}  \left[ 1 + \dfrac{\kappa_\rho m}{2M}\right]^2 \dfrac{1}{m  - 2M}  \left(1 + 2 \vec{t_\rho} \cdot \vec{t_N} \right) \left(1 + 2 \vec{s_\rho} \cdot \vec{s_N} \right).
\end{eqnarray}
Once again, using the KSRF relation for the coupling constant $g$ and pion decay constant, and recalling the fact that the masses
of the mesons are very similar to those of baryons in the present case, we can see that the contribution of the  u-channel diagram
is not negligibly small as compared to the t-channel vector meson exchange or the contact term of the HLS Lagrangian. Moreover, the contact interaction and the u-channel diagram also lead to spin-spin interactions which can play an important role in understanding of the resonances generated in these systems. However it should be mentioned that the s-channel interaction is found to be relatively weak but its contribution for other SU(3) channels needs to be checked. 

Before going ahead, we would like to remark here that we have neglected possible momentum dependence in deriving  all the interactions in our work. Such an approximation is suitable to the studies of loosely bound systems. As a consequence, the interactions are like delta functions in the coordinate space
and are  treated as separable ones while solving the scattering equations.

\subsection{Generalization to SU(3)}

Certainly,  the coupled channel effect in the SU(3) systems can be very important.  Hence, we generalize the interactions discussed in the previous section to SU(3).

In this work we study the systems with total strangeness zero.
The relevant vector meson-baryon channels  in SU(3) are: $\rho N$, $\omega N$, $\phi N$, $K^* \Lambda$ and $K^* \Sigma$.

The SU(3) generalized form of the Lagrangian for the vector-baryon interaction given by Eq.~(\ref{rhoNL}) is obtained, following \cite{jenkins, meissner, jido}, as
\begin{eqnarray} \label{vbb}
\mathcal{L}_{VB}&=& -g \Biggl\{ \langle \bar{B} \gamma_\mu \left[ V^\mu, B \right] \rangle + \langle \bar{B} \gamma_\mu B \rangle  \langle  V^\mu \rangle  
\Biggr. \\ \nonumber
&+&\left. \frac{1}{4 M} \left( F \langle \bar{B} \sigma_{\mu\nu} \left[ V^{\mu\nu}, B \right] \rangle  + D \langle \bar{B} \sigma_{\mu\nu} \left\{ V^{\mu\nu}, B \right\} \rangle\right)\right\},
\end{eqnarray}
where the  tensor field of the vector mesons is given by
\begin{equation}
V^{\mu\nu} = \partial^{\mu} V^\nu - \partial^{\nu} V^\mu + ig \left[V^\mu, V^\nu \right], \label{tensor}
\end{equation}
and the constant $D$ = 2.4 and $F$ = 0.82. These values were found to well reproduce the magnetic moments of the baryons in Ref.~\cite{jido}.
Further, in our normalization scheme,
\begin{eqnarray}
V =\frac{1}{2}
\left( \begin{array}{ccc}
\rho^0 + \omega & \sqrt{2}\rho^+ & \sqrt{2}K^{*^+}\\
&& \\
\sqrt{2}\rho^-& -\rho^0 + \omega & \sqrt{2}K^{*^0}\\
&&\\
\sqrt{2}K^{*^-} &\sqrt{2}\bar{K}^{*^0} & \sqrt{2} \phi 
\end{array}\right)
\end{eqnarray}
and
\begin{eqnarray}
B =
\left( \begin{array}{ccc}
 \frac{1}{\sqrt{6}} \Lambda + \frac{1}{\sqrt{2}} \Sigma^0& \Sigma^+ & p\\
&& \\
\Sigma^-&\frac{1}{\sqrt{6}} \Lambda- \frac{1}{\sqrt{2}} \Sigma^0 &n\\
&&\\
\Xi^- &\Xi^0 & -\sqrt{\frac{2}{3}} \Lambda 
\end{array}\right).
\end{eqnarray}

However, to obtain the right couplings for the physical $\omega$ and $\phi$ meson  at  the meson-baryon-baryon vertices, we need to consider the mixing of their octet and singlet components. Under the ideal mixing assumption, we write 
\begin{eqnarray}\nonumber
\omega &=& \sqrt{\dfrac{1}{3}} \, \omega_8 + \sqrt{\dfrac{2}{3}} \,\omega_0 \\
\phi &=&  -\sqrt{\dfrac{2}{3}}  \,\phi_8 + \sqrt{\dfrac{1}{3}} \,\phi_0,
\end{eqnarray}
and use only the octet part of these wave function in Eq.~(\ref{vbb}). In other words, the Lagrangian given by Eq.~(\ref{vbb})
corresponds to the interaction between octet vector mesons and octet baryons. For the singlet states we have
\begin{eqnarray} \label{singlet}
\mathcal{L}_{V_0BB} = -g \Biggl\{  \langle \bar{B} \gamma_\mu B \rangle  \langle  V_0^\mu \rangle  
+ \frac{ C_0}{4 M}  \langle \bar{B} \sigma_{\mu\nu}  V_0^{\mu\nu} B  \rangle  \Biggl\},
\end{eqnarray}
where the constant $C_0$ is chosen to be $3F - D$ such that  the $\phi NN$ vertex is null and  the anomalous magnetic 
coupling of the $\omega NN$ vertex gives $\kappa_\omega \simeq 3F - D$. These results, together with the anomalous magnetic coupling at the
$\rho NN$ vertex, which is $D + F = \kappa_\rho$, lead  to a consistent formalism.

Thus, in general,  the Lagrangian for the Yukawa type vertices, needed for s-, t-, and u-channel diagrams, is explicitly written as
\begin{eqnarray} \nonumber
\mathcal{L}_{VBB}&=& -g \biggl\{ \langle \bar{B} \gamma_\mu \left[ V_8^\mu, B \right] \rangle + \frac{1}{4 M} \Bigl( F \langle \bar{B} \sigma_{\mu\nu} \left[ \partial^{\mu} V_8^\nu - \partial^{\nu} V_8^\mu, B \right] \rangle \Bigr.
\biggr.   \\ \nonumber
&+&\Bigl.  D \langle \bar{B} \sigma_{\mu\nu} \left\{ \partial^{\mu} V_8^\nu - \partial^{\nu} V_8^\mu, B \right\} \rangle\Bigr)
+ \langle \bar{B} \gamma_\mu B \rangle  \langle  V_0^\mu \rangle  \\
&+&\left.  \frac{ C_0}{4 M}  \langle \bar{B} \sigma_{\mu\nu}  V_0^{\mu\nu} B  \rangle\right\}\label{yukawaL}
\end{eqnarray}
and  the term related to the two vector fields of  Eq.(\ref{tensor}) leads to a two meson-two baryon contact  interaction when plugged in Eq.(\ref{vbb}), which is trivially null for singlet meson-baryon interaction, thus giving
\begin{equation}
\mathcal{L}_{VVBB} = - \frac{g}{4 M} \Bigl\{ F \langle \bar{B} \sigma_{\mu\nu} \left[ ig \left[V_8^\mu, V_8^\nu \right], B \right] \rangle  + D \langle \bar{B} \sigma_{\mu\nu} \left\{  ig \left[V_8^\mu, V_8^\nu \right], B \right\} \rangle \Bigr\}. \label{contactl}
\end{equation}

Using the kinetic term of the hidden local symmetry Lagrangian in SU(3) for the three-vector meson vertices, we have
\begin{equation}
\mathcal{L}_{3V} \in - \frac{1}{2} \langle V^{\mu\nu} V_{\mu\nu} \rangle
\end{equation}
which, in conjunction with  Eq.~(\ref{yukawaL}), gives the t-channel interactions which are in agreement with those given in Ref.~\cite{eulogio}.
Thus,  the corresponding $T$-matrices at the lowest order are 
\begin{equation} \label{vwt}
V^{t}_{ij} = - C^{t}_{ij} \frac{1}{4 f_\pi^2} (K_1^0 + K_2^0) \vec{\epsilon_1}.\vec{\epsilon_2},
\end{equation}
which are  scalars in the spin space. The coefficients $C^{t}_{ij}$ are the same as those obtained in Ref.~\cite{eulogio}.  

Next we obtain the contact interaction in SU(3), using the Lagrangian of Eq.(\ref{contactl}) which
leads to a spin dependent $T$-matrix as discussed in the previous sub-section
\begin{equation}\label{vcontact}
V^{C.T}_{ij} = i C^{C.T}_{ij} \frac{g^2}{2 M} \vec{\sigma}\cdot\vec{\epsilon_2} \times \vec{\epsilon_1}.
\end{equation}
It is interesting to re-emphasize here on the spin structure of the contact interaction which contains $\vec{\sigma}\cdot\vec{\epsilon_2} \times \vec{\epsilon_1}$, where   $\vec{\epsilon_2} \times \vec{\epsilon_1}$ works as a spin operator  which acts on spin one particles. Hence  $\vec{\sigma}\cdot\vec{\epsilon_2} \times \vec{\epsilon_1}$ is equivalent to  $2i \vec{s} \cdot \vec{S}$, a spin-spin interaction, where $\vec{s} \, (\vec{S})$ denote the spin half (integral spin one) operator. It can be easily seen then that Eq.~(\ref{vcontact}) leads to 
\begin{eqnarray}\label{vcontact1}
V^{C.T}_{ij} &=& \,\,\,\,\,C^{C.T}_{ij} \, \,\frac{g^2}{ M} \,\,\,\, {\rm\,\, for\,\, {\it s} = 1/2,} \\
\label{vcontact3}
V^{C.T}_{ij} &=& - C^{C.T}_{ij} \frac{g^2}{ 2 M} \,\,\,\, {\rm\,\, for\,\, {\it s} = 3/2.}
\end{eqnarray}
The coefficients $C^{C.T}_{ij}$ for these potentials, projected on the ispospin 1/2 and 3/2 bases are listed in Tables~\ref{contact_iso_halfb} 
\begin{table}[htbp]
\caption{ $C^{C.T}_{ij}$ coefficients  for the contact interaction in the isospin 1/2 base.}\label{contact_iso_halfb}
\begin{ruledtabular}
\begin{tabular}{cccccc}
&&&&&\\
$C^{C.T}_{ij}$ & $\rho N$& $\omega N$ & $\phi N$& $K^* \Lambda$  & $K^* \Sigma$\\
&&&&&\\
\hline
&&&&&\\
$\rho N$ & (D+F) & 0 & 0 &  $\dfrac {(D+3F)}{4}$ & $-\dfrac{(F-D)}{4}$\\
&&&&&\\
 $\omega N$& & 0 & 0& $\dfrac{-(D+3F)}{4\sqrt{3}}$&  $-\dfrac{\sqrt{3} (F-D)}{4}$ \\
 &&&&&\\
$\phi N$& &  & 0& $\dfrac{(D+3F)}{(2\sqrt{6})}$&   $-\sqrt{\dfrac{3}{2}} \dfrac{(D-F)}{2}$ \\
&&&&&\\
$K^* \Lambda$& &  & &$\dfrac{ D}{2}$ & $-\dfrac{D}{2}$ \\
&&&&&\\
$K^*\Sigma $& & & & &$\dfrac{(2F-D)}{2}$ \\
&&&&&\\
\end{tabular}
\end{ruledtabular}
\end{table}
and \ref{contact_iso_3halfb}, respectively. From these tables, one can speculate that adding the isospin half,  spin half  contact interaction  might reduce the 
attraction given by the t-channel  and addition of isospin half,  spin 3/2 contact interaction might lead to an enhancement of attractive contribution of the t-channel. It can also be speculated that  an
opposite situation may occur in the isospin 3/2 case. 
\begin{table}[h]
\caption{ $C^{C.T}_{ij}$ coefficients for the contact interaction in the isospin 3/2 base.}\label{contact_iso_3halfb}
\begin{ruledtabular}
\begin{tabular}{ccc}
&&\\
$C^{C.T}_{ij}$ & $\rho N$&  $K^* \Sigma$\\
&&\\
\hline
&&\\
$\rho N$&$-\dfrac{(D+F)}{2}$ &$\dfrac{(D-F)}{2}$ \\
&&\\
$K^*\Sigma $& &$-\dfrac{(D+F)}{2}$ \\
&&\\
\end{tabular}
\end{ruledtabular}
\end{table}

Further, amplitudes for the diagrams involving s- and u-channel exchange have been calculated by using the Lagrangian given by Eq.~(\ref{yukawaL}) for both
the meson-baryon-baryon vertices involved. It should be mentioned that in the present formalism we consider only octet baryon exchange in these diagrams. In such a case, we obtain the following forms of the interactions in the non-relativistic approximations
\begin{eqnarray}
V^{u}_{ij}&=&C^u_{ij} \left(-\frac{g^2}{m-2 M}\right) \vec{\epsilon}_1\cdot \vec{\sigma}\,\, \vec{\epsilon}_2\cdot \vec{\sigma }, \\
V^{s}_{ij} &=&C^s_{ij} \left(\frac{g^2}{m+2 M}\right) \vec{\epsilon}_2\cdot \vec{\sigma} \,\,\vec{\epsilon}_1\cdot \vec{\sigma },
\end{eqnarray} 
where once again we have a spin-spin interaction like structure (for instance, $\vec{\epsilon }_2\cdot \vec{\sigma } \,\,\vec{\epsilon }_1\cdot \vec{\sigma }  = \vec{\epsilon}_2 \cdot \vec{\epsilon}_1 + i \vec{\sigma} \cdot  \vec{\epsilon}_2 \times \vec{\epsilon}_1$) leading to spin half contributions given by
\begin{eqnarray}
V^{u}_{ij}&=&- C^u_{ij} \left(\frac{g^2}{2 M-m}\right)  \\
V^{u}_{ij}&=&3 C^s_{ij} \left(\frac{g^2}{m+2 M}\right) 
\end{eqnarray} 
and spin 3/2 contributions 
\begin{eqnarray}
V^{u}_{ij}&=&2 C^u_{ij} \left(\frac{g^2}{2 M-m}\right)  \\
V^{s}_{ij}&=& 0.
\end{eqnarray} 

The s-channel contribution to the spin 3/2 interaction ( and also isospin 3/2 interaction) is null due to the limitation of inclusion of the octet baryon exchange in the
corresponding diagrams. The  $C^u_{ij}$ coefficients are given in   Tables. \ref{u_iso_half}
and \ref{u_iso_3half}  for the interactions projected on the isospin 1/2 and 3/2 base, respectively. The $C^s_{ij}$ coefficients for the isospin 1/2 s-channel  interaction are given in Table~\ref{s_iso_half}.

\begin{turnpage}
\begin{table}
\caption{ $C^u_{ij}$ coefficients for the  the u-channel interaction in the isospin 1/2 base, where the potential has a general form
$V^u_{ij} = C^u_{ij} \left(-\frac{g^2}{m-2 M}\right) \vec{\epsilon }_1\cdot \vec{\sigma } \vec{\epsilon }_2\cdot \vec{\sigma }$. } \label{u_iso_half}
\begin{ruledtabular}
\begin{tabular}{cccccc}
&&&&&\\
$C^u_{ij}$ & $\rho N$& $\omega N$ & $\phi N$& $K^* \Lambda$  & $K^* \Sigma$\\
&&&&&\\
\hline
&&&&&\\
$\rho N$ &$-\dfrac{\left[ (D+F)m + 2M \right]^2}{16M^2}$ & $\dfrac{\sqrt{3}}{16M^2}$& 0 & $\dfrac{Dm\left[(F-D)m+2M\right]}{8M^2}$
 & $\dfrac{\left(Fm+2M\right)\left[(F-D)m+2M\right]}{4M^2}$ \\
&&$\times \left[(D-3F)m-6M\right]$&&&\\
&&$\times \left[(D+F)m+2M\right]$&&&$\times \dfrac{Dm\left[(D+3F)m+6M\right]}{24M^2}$\\
&&&&&\\
 $\omega N$& &$\dfrac{\left[ (D-3F)m - 6M \right]^2}{16M^2}$  & 0& $-\dfrac{((3F-2D)m+6M)}{24\sqrt{3}M^2}$ &  $-\dfrac{\sqrt{3} ((F-D)m+2M)(Fm+2M)}{8M^2}$\\
  &&&&$\times ((D+3F)m+6M)$&\\
 &&&&&\\
$\phi N$& &  & 0&  $-\dfrac{((D+3F)m+6M)^2}{(24\sqrt{6} M^2)}$ & $-\sqrt{\dfrac{3}{2}} \dfrac{((F-D)m+2M)^2}{8M^2}$ \\
&&&&&\\
$K^*\Lambda $& &  & &$\dfrac{( (D-3F)m-6M)^2}{48M^2}$ & $-\dfrac{((D-3F)m-6M)}{16M^2}$ \\
& &  & & & $\times ((D+F)m+2M)$ \\
&&&&&\\
$K^* \Sigma$& & & & & $-\dfrac{((D+F)m+2M)^2}{16M^2}$ \\
&&&&&\\
\end{tabular}
\end{ruledtabular} 
\end{table}
\end{turnpage}

\begin{table}[htbp]
\caption{ $C^u_{ij}$ coefficients for the u-channel interaction in the isospin 3/2 base, where the potential has a general form
$V^u_{ij} = C^u_{ij} \left(-\frac{g^2}{m-2 M}\right) \vec{\epsilon }_1\cdot \vec{\sigma } \vec{\epsilon }_2\cdot \vec{\sigma }$. } \label{u_iso_3half}
\begin{ruledtabular}
\begin{tabular}{ccc}
&&\\
$C^u_{ij}$ & $\rho N$& $K^* \Sigma$\\
&&\\
\hline
&&\\
$\rho N$ &$\dfrac{\left[ (D+F)m + 2M \right]^2}{8M^2}$ & $\dfrac{1}{8M^2}\left\{\dfrac{-Dm}{3} \left[ \left( D+3F\right)m+6M\right]\right.$\\
&&$\left.+\left[\left(F-D\right)m+2M\right] \left[Fm+2M\right] \right\}$ \\
&&\\
$K^* \Sigma$ & & $\dfrac{((D+F)m+2M)^2}{8M^2}$ \\
&&\\
\end{tabular}
\end{ruledtabular} 
\end{table}

\begin{turnpage}
\begin{table}
\caption{ $C^s_{ij}$ coefficients for the s-channel interactions which have the general form $V^s_{ij}=C^s_{ij} \left(\frac{g^2}{m+2 M}\right) \vec{\epsilon}_2\cdot \vec{\sigma } \vec{\epsilon}_1\cdot \vec{\sigma}$.  } \label{s_iso_half}
\begin{ruledtabular}
\begin{tabular}{ccccc}
&&&&\\
$C^s_{ij}$ & $\rho N$& $\omega N$ & $K^* \Lambda$  & $K^* \Sigma$\\
&&&&\\
\hline
&&&&\\
$\rho N$ &$\dfrac{3\left[ (D+F)m - 2M \right]^2}{16M^2}$ & $\dfrac{3\sqrt{3}\left[(D+F)m-2M\right]}{8M}$&  
$\dfrac{\left((D+F)m-2M\right)\left[(D+3F)m-6M\right]}{16M^2}$ &$-\dfrac{3}{16M^2}\left[(D-F)m+2M\right]$  \\
&&&&$\times\left[(D+F)m-2M\right]$\\
&&&&\\
 $\omega N$& &$\dfrac{9}{4}$  &   $\dfrac{\sqrt{3} ((D+3F)m-6M)}{8M}$&$-\dfrac{3\sqrt{3}((D-F)m+2M)}{8M}$\\
 &&&&\\
$K^*\Lambda $& &   &$\dfrac{((D+3F)m-6M)^2}{48M^2}$&$-\dfrac{( (D+3F)m-6M)}{16 M^2}$  \\
&&&&\\
&&&&$\times ((D-F)m+2M)$\\
$K^* \Sigma$& & & &$\dfrac{3((D-F)m+2M)^2}{16M^2}$ \\
&&&&\\
\end{tabular}
\end{ruledtabular}
\end{table}
\end{turnpage}

A comment concerning the s- and u-channel diagrams is here in order.  In the present case, where we consider vector meson-baryon s-wave interaction, only the negative energy solution of the Dirac equation contributes to the s- and u-channel exchange.  The vector meson-$N$-$\bar{N}$ vertices of the resulting ``z" diagrams  are expected to be rather suppressed due to the finite structure of the hadrons. In order to take this fact into account we multiply the s- and u-channel amplitudes by the following form factor \cite{Naam,mosel1,mosel2} 
\begin{equation}
F(\Lambda, x) = \frac{\Lambda^4}{\Lambda^4 + \left( x^2 - M_x^2\right)^2},\label{formfactor}
\end{equation}  
where $x$ is the Mandelstam variable under consideration ($s$ or $u$), $M_x$ is the  mass of the baryon exchanged in such diagrams
and $\Lambda$ is a parameter which we fix as 650~MeV since it  corresponds to a reasonable average size of hadrons ($\sim$ 0.6 fm).

We have now discussed all the interactions which we will use in our study. To summarize,  vector meson-baryon interactions
have been obtained for the diagrams corresponding to the t-channel vector exchange, s- and u-channel octet baryon exchange and a contact interaction. All these interactions come from the hidden local symmetry Lagrangian and indeed we find them to be of similar order of magnitude but with different spin structure.  Hence we must consider a sum of all these diagrams, 
\begin{equation}
V^{total}_{ij}  =  V^{C.T}_{ij} + V^t_{ij} + V^u_{ij} + V^s_{ij} \label{vtot}
\end{equation}
as the minimum contribution to the lowest order vector meson-baryon interaction.  We would like to state once again that all the interactions in this work have been obtained  under the non-relativistic approximation which is adequate for the present study.  

We use $V_{total}$ of Eq.~(\ref{vtot}) as the Born term and solve Bethe Salpeter equations,
\begin{equation}
T = V + VGT,\label{bs}
\end{equation}
in the coupled channel approach, which we shall carry out in a method which is very similar to the work done in \cite{eulogio}. We do so because the results obtained in \cite{eulogio} provide us with a point of reference with which we can compare our findings. 

To solve Eq.~(\ref{bs}), we use the dimensional regularization method to calculate  the loop functions,
 \begin{eqnarray}
G (\sqrt{s}) &=& i 2 M \int \frac{d^4q}{2\pi^4} \frac{1}{(P - q)^2 - M^2 + i\epsilon} \frac{1}{q^2 - m^2 + i\epsilon} \label{loop} \\\nonumber
&=&\frac{2M}{16\pi^2} \Biggl\{ a (\mu) + \ln \frac{M^2}{\mu^2} + \frac{m^2-M^2+s}{2s}\ln \frac{m^2}{M^2} \Biggr.\\\nonumber
&+& \frac{\bar{q}}{\sqrt{s}} \Bigl[ \ln\left(s- \left( M^2 - m^2 \right) + 2\bar{q}\sqrt{s}\right) \Bigr.\\\nonumber
&+&  \ln\left(s+ \left( M^2 - m^2 \right) + 2\bar{q}\sqrt{s}\right) \\\nonumber
&-& \ln\left(-s +\left( M^2 - m^2 \right) + 2\bar{q}\sqrt{s}\right) \\\nonumber
&-& \Biggl. \Bigl. \ln\left(s- \left( M^2 - m^2 \right) + 2\bar{q}\sqrt{s}\right) \Bigr] \Biggr\},
 \end{eqnarray}
 where $\bar{q} = \lambda^{1/2} (s, M^2,m^2) /2\sqrt{s}$.  We use the subtraction constant $a(\mu) = -2$ following Ref.~\cite{eulogio} and take the pion decay constant, $f_\pi$, as 93~MeV.


 There is still an issue remaining, concerning  the width of the vector mesons, which is considerably large for the $K^*$ and $\rho$ mesons. We take care of this fact by making a convolution of  the loops over the varied mass of these mesons, again following the method used in Ref.~\cite{eulogio}, by calculating the loop as
\begin{eqnarray}\nonumber
G (\sqrt{s}) = \dfrac{1}{N} \int\limits_{(m-2\Gamma)^2}^{(m+2\Gamma)^2} d\tilde{m}^2 \left( - \dfrac{1}{\pi}\right) Im \left\{\dfrac{1}{\tilde{m}^2 - m^2 +  i m\Gamma(\tilde{m})}\right\} G(s, \tilde{m}^2, M^2),
\end{eqnarray}
where $G(s, \tilde{m}^2, M^2)$ is calculated using Eq.~(\ref{loop}) and where
\begin{eqnarray}
N = \int\limits_{(m-2\Gamma)^2}^{(m+2\Gamma)^2}  d\tilde{m}^2 \left( - \dfrac{1}{\pi}\right) Im \left\{\dfrac{1}{\tilde{m}^2 - m^2 + i m \Gamma(\tilde{m})}\right\},
\end{eqnarray}
with
\begin{equation}
\Gamma(\tilde{m}) = \Gamma_{meson} \left(\dfrac{m^2}{\tilde{m}^2}\right)  \left(\dfrac{\lambda^{1/2}(\tilde{m}^2, m_d^2, m_d^{\prime\, 2})/2\tilde{m}} {\lambda^{1/2}(m^2, m_d^2, m_d^{\prime\, 2})/2m}\right)^3 \theta \left( \tilde{m} - m_d - m_d^\prime \right).
\end{equation}
In the above equation, $m_d, m_d^\prime$ denote the masses of the decay products of the vector mesons,  i.e., pion masses in case of $\rho$, and
kaon and pion mass in case of $K^*$.  We use $\Gamma_\rho$ and $\Gamma_{K^*} $ as 149.4~MeV and 50.5~MeV, respectively.


\section{Results and discussion}
In the previous section we have discussed the interaction of the vector-meson with the octet baryon obtained  from four different diagrams corresponding to the  t-, s-, u-channel exchange and a contact term. We have tabulated the lowest order amplitudes
obtained from these diagrams for the vector meson-octet baryon channels  having a total charge and strangeness zero. Using these contributions we have solved Bethe-Salpeter equations in the coupled channel approach and in this section we will discuss the results of our calculations.

\subsection{t-channel exchange}
\begin{figure}[h]
\begin{center}
\includegraphics[width=10cm,height=8cm]{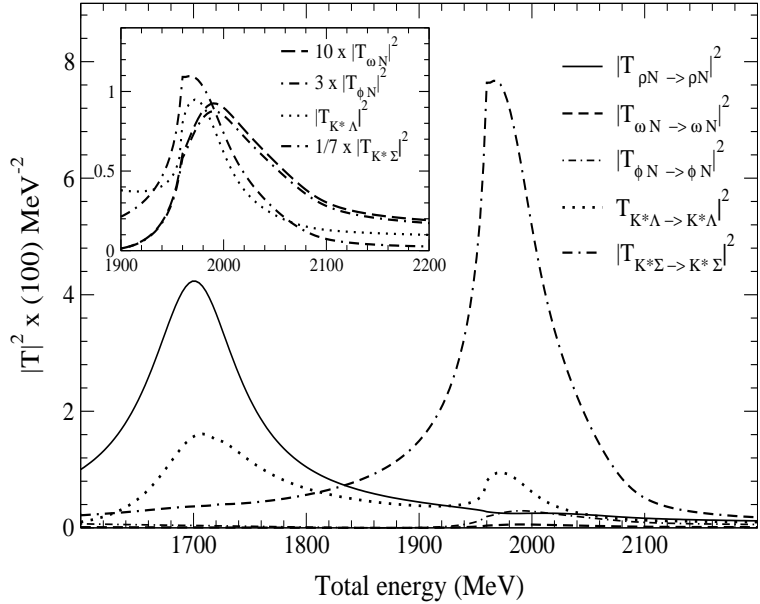}
\caption{   Squared $T$-matrices obtained by solving the Bethe Salpeter equation using t-channel interaction. Two clear peaks can be seen in these  amplitudes around 1700~MeV and 2000~MeV. The inset figure shows the same amplitudes multiplied by arbitrary factors in the 2~GeV energy region. The purpose of the inset figure is to show that the peak in the 2~GeV region in the  amplitudes for the $\omega N$ and $\phi N$ channels is slightly shifted as compared to the one in the  $K^* \Lambda$ and $K^* \Sigma$ channels. }
\label{tallch}
\end{center}
\end{figure}

We will first discuss the results of the calculations done by using the t-channel interaction only. The vector meson-baryon system can have total isospin and spin  1/2 or 3/2. However, as it was shown in Ref.~\cite{eulogio} and as we have already mentioned in the previous section, the structure of the interaction obtained from the t-channel diagrams is spin independent. Further, the interaction in the isospin 3/2 case is repulsive and, as expected, it does not result in generation of  any state. 

The isospin 1/2 interaction does result in finding of some states, which are spin degenerate in nature.
We present the squared amplitudes, $|T|^2$, obtained in this case, in Fig.~\ref{tallch}. As can be seen in this (main) figure, very much in agreement with the work of Ref.~\cite{eulogio}, we find two peaks in the squared amplitude: one around 1700~MeV and another around 2~GeV.  These two peaks can be identified as $N^*$ 's with mass around 1700 and 2000~MeV, with spin-parity $1/2^-$ and $3/2^-$ since the calculations have been done in s-wave and the basic interaction is spin degenerate. There do exist $N^*$ resonances with such properties \cite{pdg} and relating them to the peaks shown in Fig.~\ref{tallch} seems reasonable as suggested in Ref.~\cite{eulogio}. However we would not yet make any such analysis since we expect these results to get modified by addition of more interactions.

\begin{figure}[ht]
\begin{center}
\includegraphics[width=15cm,height=12cm,angle=0]{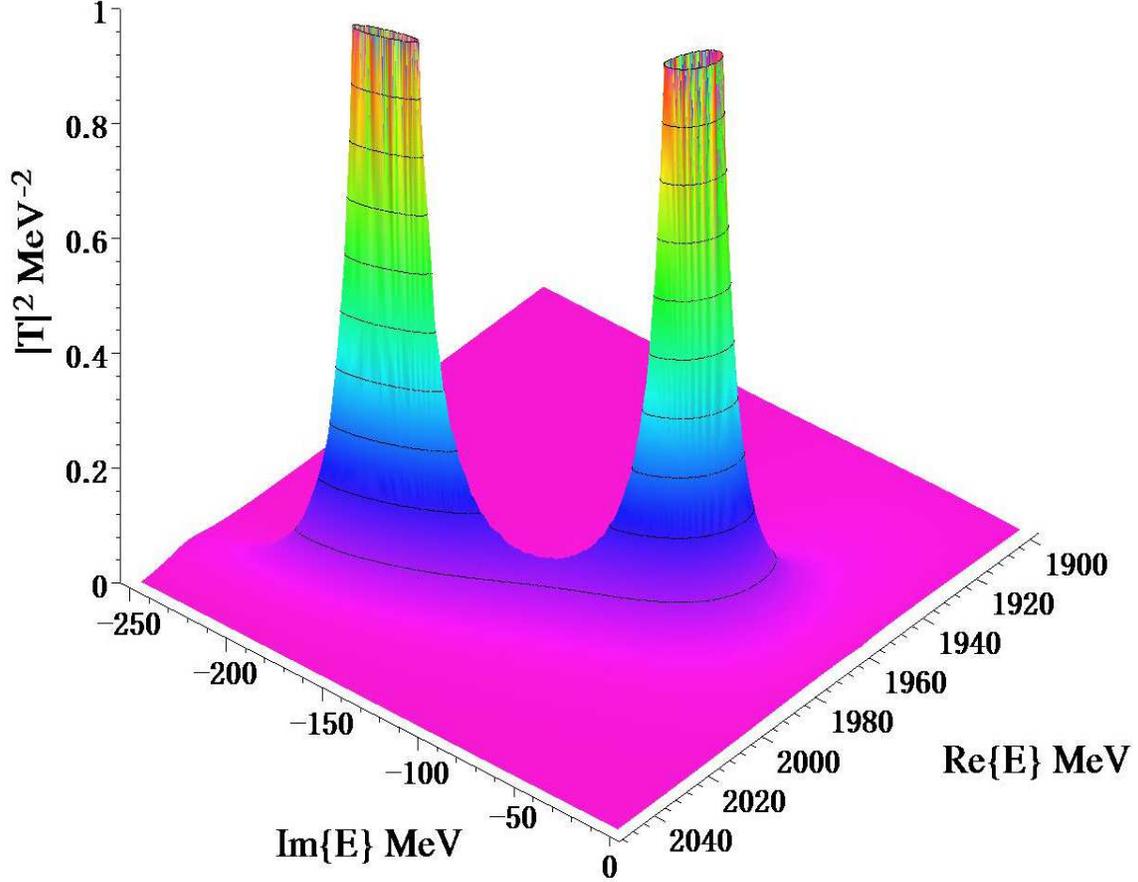}
\caption{   Two pole structure of a possible $N^*$ with mass $\sim$ 2000~MeV,  spin parity $1/2^-$ and $3/2^-$.}
\label{twopoles}
\end{center}
\end{figure} 

We have also  searched for poles in the complex plane for the amplitudes depicted in Fig.~\ref{tallch}. We find a bound state pole corresponding to the lower energy peak,  on the real axis, at $1702 - i0$~MeV, when the width of the $\rho$ and $K^*$ mesons is not taken into account in the calculations. 
A consideration of the mass distributions of these mesons makes it difficult to look for poles, especially in the energy region close to a threshold as is the case of this pole. This problem was already discussed in  Ref.~\cite{eulogio}.  This pole seems to couple strongly to the $\rho N$ channel as can be seen in the Fig.~\ref{tallch}
and as the couplings calculated in Ref.~\cite{eulogio} indicate.

 In case of the peak seen around the 2~GeV region, our searching for a pole gives results which are  slightly different to the ones discussed in Ref.~\cite{eulogio}. We find a double pole structure for the peak around 2~GeV as shown in Fig.~\ref{twopoles}. 
The pole positions obtained without convoluting the loops are: $1974 - i44$~MeV and $2051 - i153$~MeV and those obtained on convolution of the loops are $1980 - i58$~MeV and $2019 - i164$~MeV. We find the pole  at higher energy to couple strongly to the $K^* \Sigma$ channel, so much that it can be generated even by making a single $K^* \Sigma$ channel calculation at $2072 - i0$~MeV (without convoluting the loops), which shifts to $2082 - i25$~MeV on adding the $\phi N$ and $K^* \Lambda$ coupled channels and finally to $2051 - i164$~MeV on adding $\rho N$ and $\omega N$. The other pole, which is found near  $1970$~MeV,  gets generated due to the coupled channel dynamics  of  $\phi N$ and $K^* \Lambda$. It is interesting that the diagonal interactions for these two channels are null in this case, which makes the coupled channel effect  imperative for the generation of this pole.  

 These two poles seem  to couple more to the $\phi N$, $ K^* \Lambda$ and $K^* \Sigma$ channels. We will verify this later in the discussions on the couplings calculated for different poles.  We have tried to show the effect of this two pole structure on the real energy axis in the inset picture of Fig.~\ref{tallch} where
 we have multiplied arbitrary factors to the   $\omega N, \phi N, K^* \Lambda, K^* \Sigma$ diagonal $T$-matrices to compare the peak positions in these  amplitudes around 2~GeV.  It could be argued that one out of these two poles is too wide and may not be important in the sense that it's effect cannot be  easily seen in real experimental data. Such a consequence might be close to reality, however, the picture will change on addition of more interactions as we shall see in the following sections, and this two pole structure will play a crucial role in understanding the results.

\subsection{Addition of more diagrams}
The t-channel interaction leads to generation of two pairs of spin 1/2-3/2 degenerate resonances with isospin 1/2. These resonances have a mass
$\sim$ 1700~MeV and $\sim$ 2000~MeV with the latter one possessing a two pole structure, one of which could be interpreted as a $K^* \Sigma$ bound state
and the other could be interpreted as a $K^* \Lambda - \phi N$ resonance. 
We will now discuss the results obtained by adding to the t-channel interactions, the contact term, the s- and u-channel, where the resultant interaction has a $\vec{s}\cdot \vec{S}$ structure (spin-spin interaction).  We thus expect to lift the spin degeneracy of the states obtained by considering the t-channel
diagrams as the tree-level amplitudes. Let us first consider the case of total isospin 1/2  of the meson-baryon systems.

\subsubsection{Isospin=1/2}
\begin{figure}[ht]
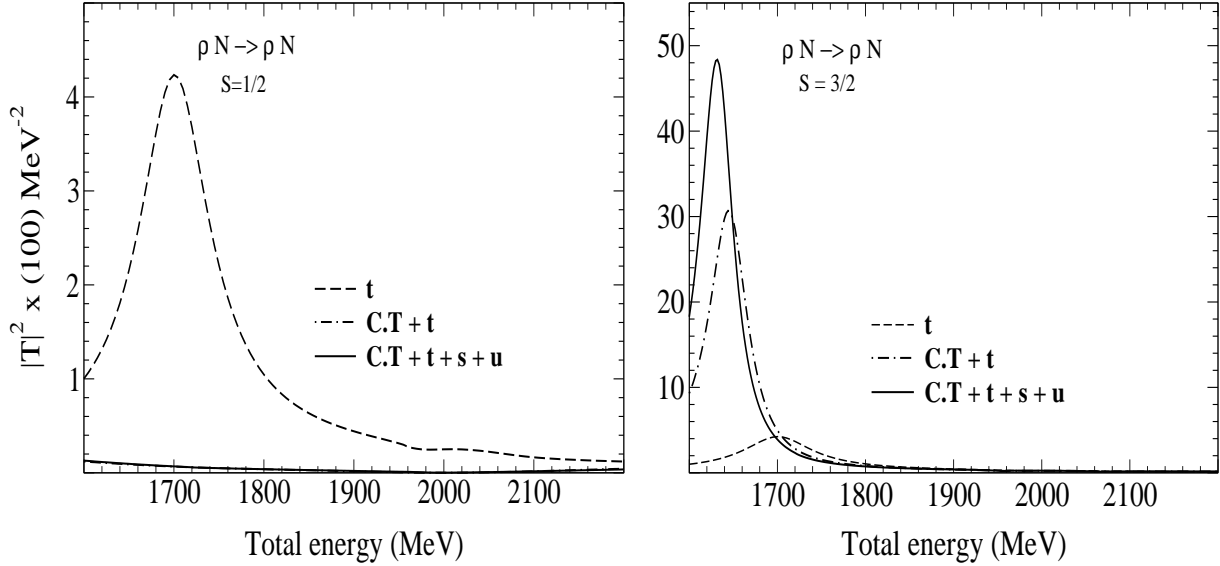

\begin{minipage}[b]{0.495\linewidth}
\includegraphics[width=8.2cm,height=7.5cm]{rhon_all_int_s_half.eps}
\end{minipage}
\begin{minipage}[b]{0.495\linewidth}
\includegraphics[width=7.5cm,height=7.5cm]{rhon_all_int_s_3half.eps}
\end{minipage}
\caption{  Squared amplitude, $|T|^2$ of the process $\rho N \rightarrow \rho N$ as a function of the total energy: for spin 1/2 depicted in the left figure
and spin 3/2 in the right one.  The labels s, t, u, C.T refer to the calculations done by taking  s-, t-, u-channel, contact term, respectively into account.  The solid lines in these figures correspond to the calculations done by taking a sum of all these diagrams,  as explained in the text.  The dashed lines are the results obtained by considering t-channel and the dash-dotted lines show the 
effect of addition of the contact term.}\label{rhoN}
\end{figure}

In Fig.~\ref{rhoN} we show the squared amplitude, $|T|^2$, for the  $\rho N $ channel , for the spin half (left panel) and spin three-half (right panel) case.  The result obtained
by considering  the t-channel interaction alone is shown by a dashed line (which is same as the
solid line of Fig.~\ref{tallch}).  The result of adding the contact interaction  is shown by  a dash-dotted line, and that obtained by further adding the s- and u-channel  exchange  is shown by a solid line. All these calculations have been carried out by using the procedure of convolution of the loops to take in to account the fact that the $\rho$ and $K^*$ mesons possess a non-negligible width. Henceforth we shall stick to the discussion of the results obtained by convoluting the loops unless otherwise is stated.  

It can be seen that the calculations done by adding the contact term to the t-channel interaction give very different results (dash-dotted lines) as compared to the ones obtained by considering the t-channel alone (dashed lines). The left panel shows that the clear peak  found
in spin degenerate t-channel calculations disappears on adding the spin half contact interaction. The magnitude of the squared amplitude falls by about 
two orders of magnitude in the 1700~MeV region  and no structure is found on adding the contact term, $V_{C.T}$.

The results are quite different for the spin 3/2 case as shown in the adjacent figure (right panel).  In this case the peak obtained from t-channel calculations gets enhanced by a factor 10 but shifts by about 50~MeV to the lower energies.  These results can be very well understood by looking at the structure of the contact interaction given by   Eqs. (\ref{vcontact1}) and (\ref{vcontact3}) along with the coefficients $C^{C.T}_{ij}$ of Table~\ref{contact_iso_halfb}.  It can be seen that the contact term is repulsive  for most channels in spin half case but is mostly  attractive  in the spin 3/2 case, with the order of magnitude being similar to the t-channel interaction always. Thus the spin 1/2 contact term  reduces the attractive t-channel interaction  in the vector-baryon system but adds to the attraction in spin 3/2 case. As a result  the spin degenerate peak at 1700~MeV in the amplitudes obtained by t-channel interaction disappears in spin 1/2 case, and the spin 3/2 results show the  peak shifted to lower energies with much larger magnitude. Further addition of the s- and u-channel diagrams enhances the effect produced by the contact term in the $T$-matrices as shown by the solid lines.

\begin{figure}[ht]
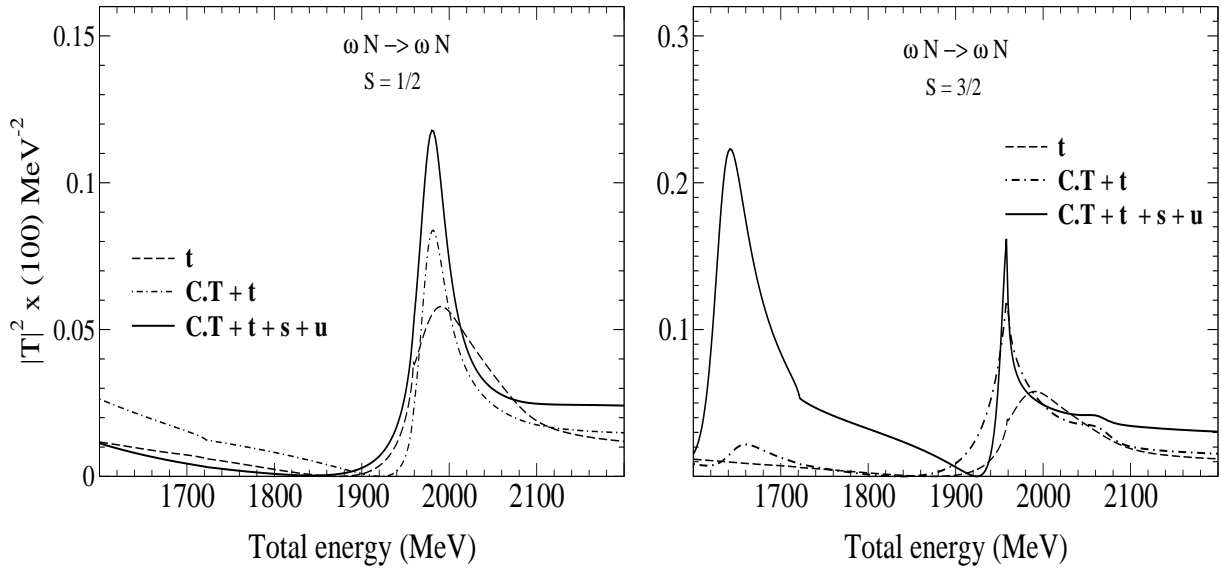

\begin{minipage}[b]{0.495\linewidth}
\centering
\includegraphics[width=8.2cm,height=7.5cm]{omegan_all_int_s_half.eps}
\end{minipage}
\begin{minipage}[b]{0.495\linewidth}
\centering
\includegraphics[width=7.5cm,height=7.5cm]{omegan_all_int_s_3half.eps}
\end{minipage}
\caption{   Squared amplitude of the process $\omega N \rightarrow \omega N$ as a function of the total energy. The meaning of the panels, labels and the lines continues to be same as in Fig.~\ref{rhoN}. }\label{omegaN}
\end{figure}


Next, we show the  squared amplitude for  the $\omega N$ channel in Fig.~\ref{omegaN}.  Here we see that the contact term added to the t-channel diagram  produces a slightly enhanced peak structure in the spin half amplitude, but a factor three larger matrix element, in the peak region, in case of spin 3/2 which shows a pronounced peak at slightly lower energies ($\sim$1960~MeV) and a bump at $\sim$1660~MeV.  In this case, further addition of the s- and u-channel diagrams 
gives rise to yet another clear peak in spin 3/2 amplitude, at about 1650~MeV.  Thus,  our total  $\omega N$ amplitude shows two  peaks in the spin 3/2
case. Another interesting feature seen in the spin 3/2 $T$-matrices is a broad bump near 2050~MeV. 
\begin{figure}[ht]
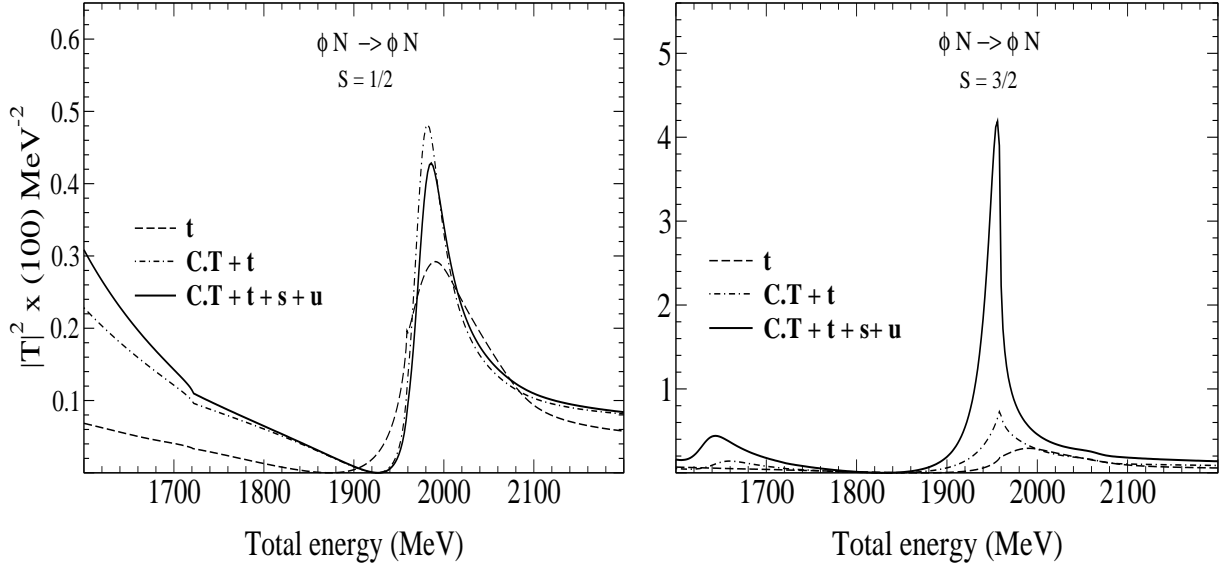

\begin{minipage}[b]{0.495\linewidth}
\centering
\includegraphics[width=8.2cm,height=7.5cm]{phin_all_int_s_half.eps}
\end{minipage}
\begin{minipage}[b]{0.495\linewidth}
\centering
\includegraphics[width=7.5cm,height=7.5cm]{phin_all_int_s_3half.eps}
\end{minipage}
\caption{ Squared amplitudes for  spin 1/2 and 3/2 $\phi N \rightarrow \phi N$ process as a function of total energy. The meaning of the lines, lables and the purpose of the inset figure is same as that in  Fig.~\ref{rhoN}. }\label{phiN}
\end{figure}

Figure~\ref{phiN} depicts the squared amplitude for the $\phi N$ channel. Once again, for this channel too, the addition of the contact 
term to the t-channel interaction, produces quite some changes  in the results obtained without it.  The spin 1/2 $\phi N$ matrix element shows an enhancement
of the strength of squared matrix element, like in the $\omega N$ case. The spin 3/2 $\phi N$ amplitude shows a sharper peak around 1950~MeV by addition of the contact term, s-, and u-channel diagrams. In the total amplitude (solid line), one can also see a bump around 1650~MeV.

\begin{figure}[ht]
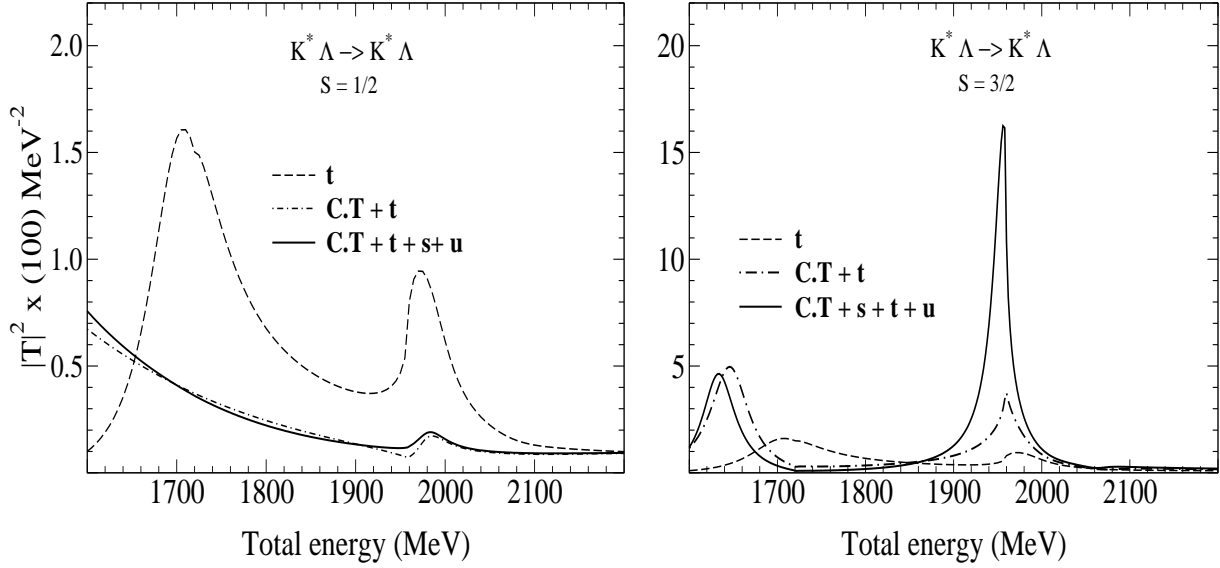

\begin{minipage}[b]{0.495\linewidth}
\centering
\includegraphics[width=8.2cm,height=7.5cm]{KstarLambda_all_int_s_half.eps}
\end{minipage}
\begin{minipage}[b]{0.495\linewidth}
\centering
\includegraphics[width=7.5cm,height=7.5cm]{KstarLambda_all_int_s_3half.eps}
\end{minipage}
\caption{   Same as the Figs.~\ref{rhoN}-\ref{phiN} but for the  $ K^* \Lambda \rightarrow K^* \Lambda$ amplitude. }\label{KstarLambda}
\end{figure}

It remains to discuss the results for  $ K^* \Lambda$ and $ K^* \Sigma$ channels. The ones corresponding to the former channel
are shown in Fig.~\ref{KstarLambda}. This is the only channel which shows two peaks in the squared amplitude calculated by taking the t-channel
interaction. Clearly, the lower energy peak disappears on adding the mostly repulsive, spin 1/2 contact interaction. The spin 3/2 case
results in two distinct  peaks (as shown by the dash-dotted line in the right panel  of Fig.~\ref{KstarLambda} ). The strength of the squared amplitude in this case gets enhanced by about a factor 5. Further addition of the s- and u-channels, although leaves the spin 1/2 amplitude almost unaltered, increases the magnitude of the spin 3/2 amplitude by another factor $\sim$ 3 near 1950~MeV. This hints towards a larger coupling of the $ K^* \Lambda$ channel to the $3/2^-$ state with mass close to 1950~MeV.  We shall verify this in the subsequent sub-section where we discuss the calculations of the couplings.

\begin{figure}[h]
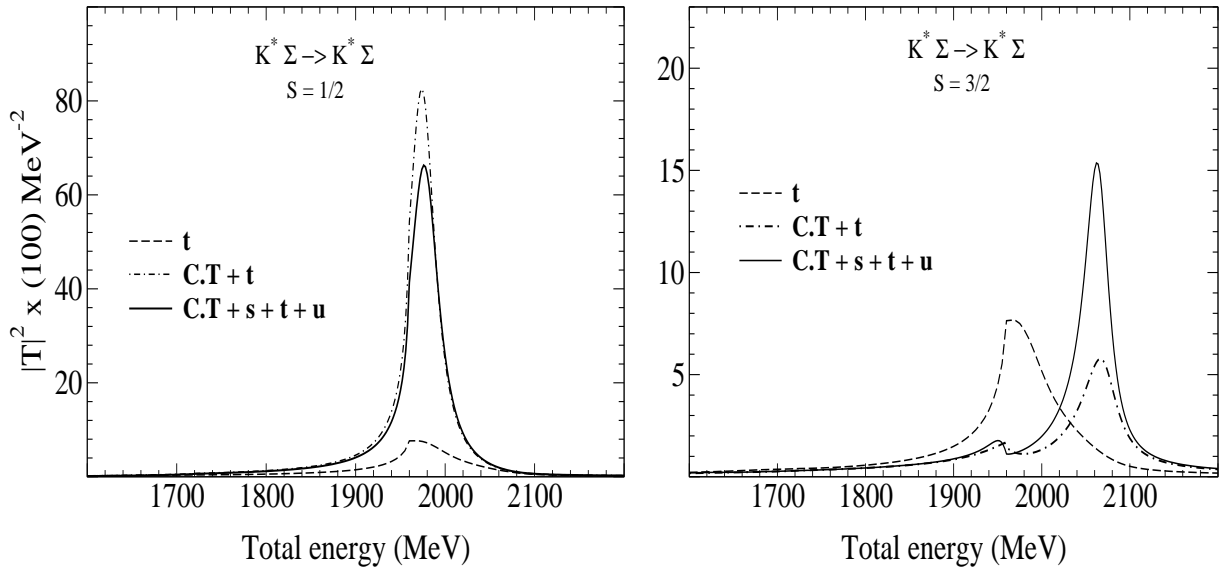

\begin{minipage}[b]{0.495\linewidth}
\centering
\includegraphics[width=8.2cm,height=7.5cm]{KstarSigma_all_int_s_half.eps}
\end{minipage}
\begin{minipage}[b]{0.495\linewidth}
\centering
\includegraphics[width=7.5cm,height=7.5cm]{KstarSigma_all_int_s_3half.eps}
\end{minipage}
\caption{   Squared amplitude for the  $ K^* \Sigma$ channel. }\label{KstarSigma}
\end{figure}

Finally, let us contemplate the $ K^* \Sigma$ coupled channel.  The squared amplitude for this channel shows  a well pronounced peak in the spin 1/2 case at $\sim$ 1975~MeV ( left panel of Fig.~\ref{KstarSigma} ).  It can be seen that addition of the contact term enhances the  strength of the peak found in the t-channel calculations (dashed line) by about an order of magnitude (dash-dotted line). The full calculations depicted as solid lines show some reduction in the strength. Nevertheless the strength of the spin 1/2 amplitude of the $K^* \Sigma$ channel remains the largest, which indicates towards its strong coupling to a possible corresponding state  with $1/2^-$  quantum number. 
The full spin 3/2 $ K^* \Sigma$ amplitude, depicted as a solid line in the right panel  of Fig.~\ref{KstarSigma}, shows a peak at  2064~MeV and a  bump around 1950~MeV. Interestingly, the spin 3/2 $K^* \Lambda$  and $\phi N$ amplitudes show a clear peak at 1950~MeV, where the $ K^* \Sigma$ amplitude shows a weak bump. 

To interpret these isospin 1/2 results  on the real axis, we have looked for poles in the complex plane. In spin 1/2 case we find a pole 
at $1977 - i22$~MeV and in the spin 3/2 case we find two poles at $1642- i0$ and  $2071- i7$~MeV.  We find that the peak structure seen in 
the $\phi N, \omega N$ and $K^* \Lambda$ amplitudes near 1960~MeV correspond to a $\phi N$ cusp for which we see a pole extremely close
to the real axis near the $\phi N$ threshold. 

To summarize  our results in the isospin 1/2 case, we study the strangeness zero vector meson-baryon system by taking a contact
term, t-, s-, and u-channel diagrams as leading order interactions. A coupled channel calculation with these interactions made on the real energy axis, leads to finding of a peak close to 1970~MeV with total spin equal to 1/2. We do not find any structure near 1700~MeV in the spin 1/2 amplitude.  In the spin 3/2 case we find a peak near 1650~MeV and  2~GeV.  We will come back to a more detailed discussion of corresponding poles found in the complex plane, in a later section.

\subsubsection{Isospin = 3/2}

\begin{figure}[h]
\begin{center}
\includegraphics[width=9cm,height=8cm]{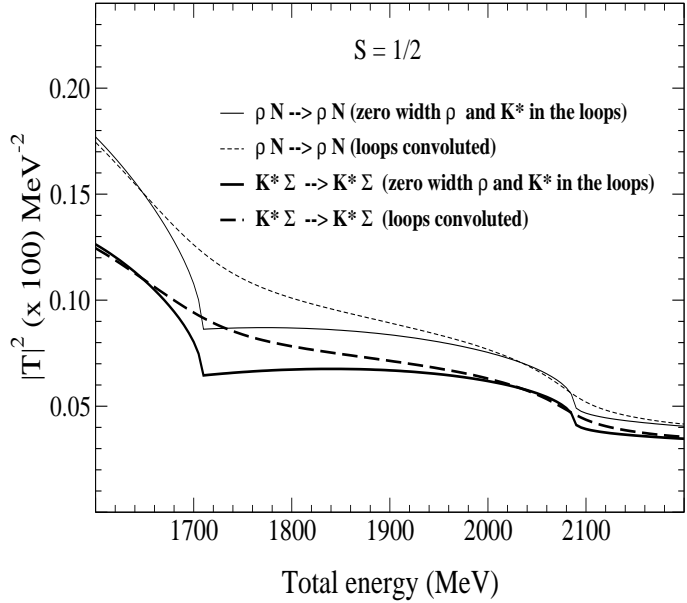}
\caption{   Isospin 3/2, spin 1/2 amplitudes obtained when only the t-channel interaction is used in the calculations. Shown in this figure are the results for the $\rho N$ and  $K^* \Sigma$ channels as thin and thick lines, respectively. }
\label{wt_iso_3half}
\end{center}
\end{figure}

In Fig.~\ref{wt_iso_3half} we show  the results of the calculation of the vector meson-baryon amplitude in the isospin 3/2, spin 1/2 space where we 
have only two coupled channels: $\rho N$ and $K^* \Sigma$. In this figure we show the results of the calculation done by considering the
t-channel interaction alone which is actually repulsive in nature. As a result, the amplitudes in this isospin are much weaker and rather flat 
as compared to the corresponding isospin half results. One can see some kinks due to opening of the channels: the thresholds of the $\rho N$ and $K^* \Sigma$ are 1709~MeV and 2085~MeV. The results for the $\rho N $ system are shown as thin solid and dashed lines and that for the  $K^* \Sigma$ are shown as thick lines. The dashed (solid) curves have been obtained by  calculating  the loops with (without) the
consideration of the widths of the $\rho$ and the $K^*$-mesons. 

\begin{figure}[ht]
\begin{center}
\includegraphics[width=9cm,height=8cm]{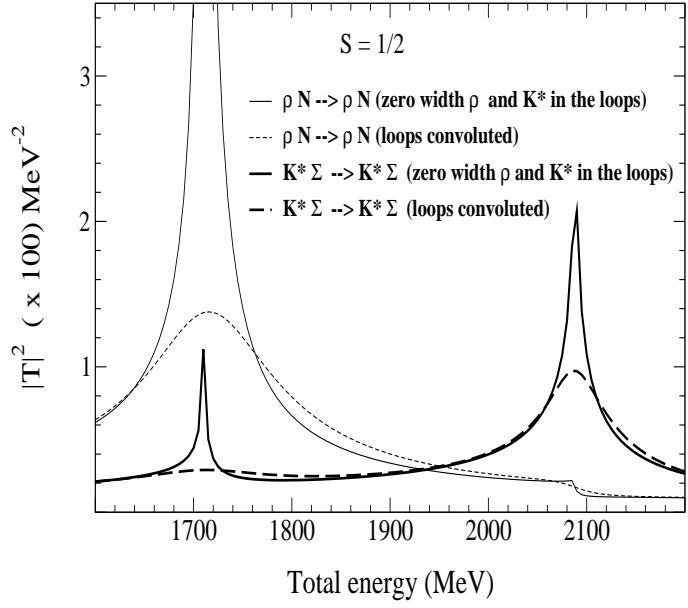}
\caption{  Results  obtained in  isospin 3/2, spin 1/2 configuration by considering the contact term of Eq.~(\ref{contactl}). }
\label{contact_iso_3half}
\end{center}
\end{figure}

Before showing the results obtained by adding more diagrams to the t-channel  in the  isospin 3/2 case, we would like
to discuss the results obtained by carrying out the calculations assuming only the contact term. The results of such a calculation are shown
in Fig.~\ref{contact_iso_3half}. The results obtained
in this case are very different to the corresponding t-channel calculations since the contact interaction is attractive in nature, in isospin 3/2 and spin 1/2 case  (see Eqs. (\ref{vcontact1}) and 
(\ref{vcontact3}) and Table~\ref{contact_iso_3halfb}). As can be seen in Fig.~\ref{contact_iso_3half}, we find
two sharp peaks: one near 1700~MeV and another near 2100~MeV. The former seems to couple strongly to the $\rho N$ channel and the latter to 
the $K^* \Sigma$ system. 
We would like to add here that although we see two peak structures in the amplitudes shown in Fig.~\ref{contact_iso_3half}, we do not find
the corresponding poles to be physical. 

Finally, let us discuss the results we obtain by adding the contact term, t- and u-channel interactions. We would
like to remark here again that in this case we have no contribution from the s-channel since in the present formalism we assume exchange of isospin 1/2 baryons only.

\begin{figure}[h!]
\begin{center}
\includegraphics[width=9cm,height=8cm]{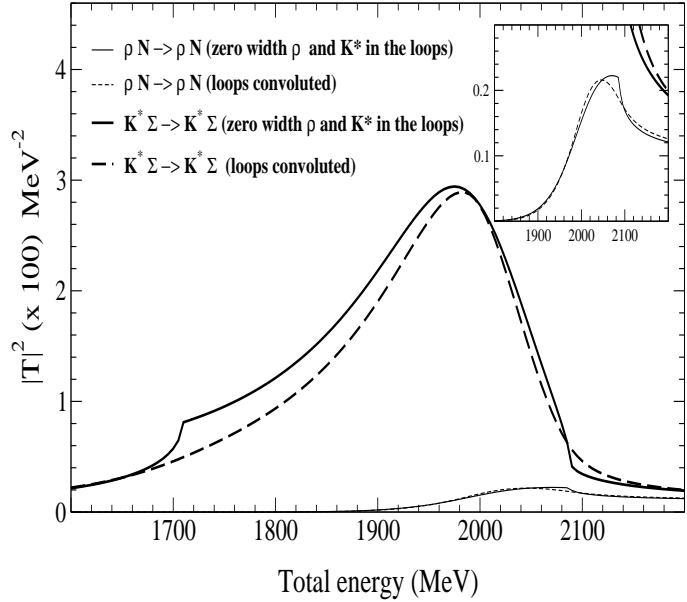}
\caption{  Squared amplitudes for isospin=3/2, spin=1/2 obtained by taking the t-channel +contact term +u-channel interaction
as the Bonn term.}
\label{all_iso_3half}
\end{center}
\end{figure}

It should be also mentioned that the inclusion of  the u-channel in the calculations does not change the results obtained by considering 
contact term and t-channel, except for a small increase in the magnitude and a  shift in the peak position by about 20~MeV. Hence we do not show the results of the calculations carried out by adding the contact term to the t-channel. We rather directly show the results obtained by adding the u-channel as well, in Fig.~\ref{all_iso_3half} . In this case, we find, one, rather broad peak around 1980~MeV which is about an order of magnitude bigger in the $K^* \Sigma$ channel.  We find a pole corresponding to it  in the second Riemann sheet at $2010 - i112$~MeV, when the loops are convoluted. A pole search  in the matrix elements calculated without convoluting the loop does not give very different results, in this case we find a pole at $2002 - i108$~MeV. 

\subsection{Poles in the complex plane and their couplings to different channels:}
In addition to finding the poles in the complex plane it is  also important to calculate the couplings of resonant/bound states to different channels, which help in understanding the behavior of the squared amplitude on the real energy axis (and hence the cross sections). 
These couplings, 
$g^{i}$,  can be  calculated by finding residues of the poles of the amplitudes, in the complex plane. However, there is one difficulty in making such a calculation in our work, that is due to the convolution of loops which makes  that a fixed threshold becomes a variable one, when the widths of the $\rho$ and $K^*$  mesons are taken into account. This difficulty was also confronted in \cite{eulogio} and the couplings were  calculated using the amplitudes on the real axis. We avoid this problem by calculating the residues of the poles in complex plane, which are found without making any convolution of the loops.  We find the couplings, $g^{i}$, in this way for all the cases. We have seen that the couplings found with or without convolution are similar, when the poles can be found in both approaches.

\begin{table}[htbp]
\caption{   Couplings  $g^{i}$  of the isospin 1/2, spin degenerate poles obtained in t-channel calculations.} \label{tcoupling}
\begin{ruledtabular}
\begin{tabular}{cccc}
$M_R - i\Gamma/2$ $\longrightarrow$ & $1702-i0$~MeV&$1974-i44$MeV &$2051-i153$~MeV\\
 ($J^\pi$) & ($1/2^-$,  $3/2^-$) &($1/2^-$, $3/2^-)$ &$(1/2^-$, $3/2^-)$ \\
channels $\downarrow$& \multicolumn{3}{c}{Couplings ($g^i$)}\\
\hline
$\rho N$&$2+i0$&$-0.1+i0.6$& $-1.9-i0.7$\\
$\omega N$&$0.1+i0$&$-1+i0.5$&  $-1.4-i0.7$\\
$\phi N$  &$-0.15+i0$&$1.4-i0.7$&$2.1+i1$\\
$K^*\Lambda$  &$1.7+i0$&$2.1+i0.8$&$1 +i1.4$\\
$K^*\Sigma$  &$-0.4+i0$&$3.7-i0.6$&$5.1+i3.2$\\
\end{tabular}
\end{ruledtabular}
\end{table}

We shall first discuss the couplings obtained for the poles found in the calculations with only  t-channel interaction. In such a calculation, as mentioned in the previous section, we find three poles corresponding to the two peaks found on the real axis.  These poles are spin degenerate in nature and possess an isospin 1/2.  The couplings obtained for these poles are listed in Table~\ref{tcoupling} which show that the pole found at $1702-i0$~MeV couples strongly to the $\rho N$ and $K^*\Lambda$ channels and indeed that is what we also find from the behavior of the amplitudes calculated on the real axis (shown in Fig.~\ref{twopoles}). Both the other  two poles listed in Table~\ref{tcoupling}: $1974-i44$ and $2051-i153$~MeV, the latter of which was not discussed in Ref.~\cite{eulogio}, seem to couple strongly to $\phi N$, $K^*\Lambda$ and $K^*\Sigma$. 

Next, let us discuss the couplings of the poles found in the calculations done with a sum of contact, t-, s-, and, u-channel interactions. As we have already seen, such a calculation lifts the spin degeneracy of the results found in the t-channel calculations. In Table~\ref{isohalf_couplings} we show the poles found in the isospin 1/2 case. Once again, we find three poles but two with spin parity $3/2^-$: at  $1642- i0 $, $2001- i7$~MeV and one with $1/2^-$ at $1977 - i22$~MeV (in the amplitude calculated without convoluting the loops). 
\begin{table}[htbp]
\caption{Couplings  $g^{i}$  for the isospin 1/2 poles obtained in a calculation done with all the four interactions shown in Fig.~\ref{diagram}.} \label{isohalf_couplings}
\begin{ruledtabular}
\begin{tabular}{cccc}
$M_R - i\Gamma/2$  $\longrightarrow$  & $1642- i0 $~MeV & $2071- i7$~MeV&$1977 - i22$~MeV \\
 ($J^\pi$) &($3/2^-$) &$(3/2^-$)&($1/2^-$) \\
channels $\downarrow$& \multicolumn{3}{c}{Couplings ($g^i$)}\\
\hline
$\rho N$&$3.9+i0$&$0.02-i0.4$&$0.04+i0.2$\\
$\omega N$&$0.9+i0$&$-0.1-i0.1$&$-0.7+i0.2$\\
$\phi N$  &$-0.8+i0$&$0.14 + i0.2$&$1.1-i0.5$\\
$K^*\Lambda$  &$2.2+i0$&$-0.3+i0.35$&$0.6+i0.08$\\
$K^*\Sigma$ &$-0.3+i0$ &$2.4+i0.3$ &$4.4 - i0.1$\\
\end{tabular}
\end{ruledtabular}
\end{table}

The spin 3/2 pole found at $1642- i0 $~MeV, which acquires a width when the loops are convoluted and is found at $1637 - i35$ MeV, seems to couple strongly to the  $\rho N$ and $K^*\Lambda$ channels just like in the t-channel case.  However we do not find a spin 1/2 pole in this energy region. We relate our spin 3/2 pole at $1637 - i35 $MeV with $N^* (1700)~D_{13}$, for which 
the particle data group (PDG) \cite{pdg} lists a branching ratio of about 35$\%$ to the $\rho N$ channel and only 5-15 $\%$ to $\pi N$. Our results are  consistent with the findings of a study made by the Juelich group 
\cite{michael}, where  $N^* (1700)~D_{13}$ has been found to get dynamically generated basically by the  $\rho N$ interaction  and a very small coupling to the  $\pi N$  channel has been found.  

The other $3/2^-$ pole found in our study at $2071- i7$~MeV is basically a $K^*\Sigma$ bound state. The reason for its small width is its strong coupling to $K^*\Sigma$ which is a closed channel (while neglecting the $K^*$ width) at this energy  and its very weak couplings to all other channels. However, when the loops are convoluted, we find this pole to shift to $2071 -i70$ MeV.  There is a resonance listed by the PDG with mass and spin parity strikingly similar to this resonance: the  $N^* (2080)~D_{13}$ for which a branching ratio of 21$\%$ is listed for the $\omega N$ decay channel.  This is indeed in agreement with our findings since, although the coupling of this pole
to $\omega N, \phi N, K^* \Lambda$ are similar, $\omega N$ offers much larger phase space for the decay of the resonance. Also, the several hundred MeV width of $N^* (2080)~D_{13}$  listed in \cite{pdg} in comparison with our results is understandable since we have not taken pseudoscalar meson-octet baryon or pseudoscalar-decuplet baryon channels  into account which provide much larger phase space for the decay.  We thus identify the spin 3/2 resonance  found at  $2071- i70$~MeV with  $N^* (2080)~D_{13}$.

Finally, we find only one resonance with spin parity $1/2^-$ at $1977 - i22$~MeV which couples strongly to $K^*\Sigma$ and $\phi N$  channels.  However its properties can only be studied in the latter one since it is open at these energies.  This pole appears at the same position even when the convolution of the loops is made. It could be identified with  $N^*(2090)~S_{11}$ although the status of this resonance is rather poor as found by the PDG \cite{pdg}. However, there is not much 
information available on the analysis of  the  $\phi N$  channel near 2~GeV, in the partial wave $S_{11}$, which, according to our study, might improve the status of the $N^*(2090)~S_{11}$ resonance. Indeed an enhancement of the cross section is seen near 2~GeV region, in the photo-production of the $\phi$ meson on a nucleon studied by the LEPS group \cite{leps}. This enhancement could possibly be explained in terms of the spin $1/2^-$ resonance found in our work. In fact a phenomenological analysis of the $\gamma N \rightarrow \phi N$ reaction  showed that a better fit to the LEPS data \cite{leps} was found
when a $J^\pi = 1/2^-$ resonance was included in their study as compared to the one obtained by including a $1/2^+$  resonance \cite{ozakisan}.  It should be interesting to clarify this with more detail  in future.

\begin{table}[htbp]
\caption{   Same as Table~\ref{isohalf_couplings}   but for the case of isospin 3/2, spin 1/2. } \label{iso3half_couplings}
\begin{ruledtabular}
\begin{tabular}{ccc}
$M_R - i\Gamma/2$ $\longrightarrow$  &$2006 - i112 $~MeV &$2002 -i108$~MeV \\
 &(with conv)& (without conv)\\
channels $\downarrow$& \multicolumn{2}{c}{Couplings ($g^i$)}\\
\hline
$\rho N$&$-1.6+i1.4$& $-1.7 + i1.2$\\
$K^*\Sigma$ &$4.5+i0.7$&$4.5+i1.3$ \\
\end{tabular}
\end{ruledtabular}
\end{table}
Finally we discuss the couplings of the pole found in the isospin 3/2. We find a pole in  spin1/2 case, in 2~GeV region with a width of about 100~MeV. We show the precise pole positions and couplings found for this pole, in the calculations done with and without convoluted loops,  in Table~\ref{iso3half_couplings}.  There are only two coupled channels contributing to this isospin and the resulting pole is found to couple strongly to the $K^*\Sigma$ channel. We relate this state with $\Delta (1900)~S_{31}$.

\subsection{Pole Flow}

It is intriguing that we started calculating vector meson-baryon amplitudes by taking a t-channel interaction  and we found six poles in isospin 1/2, which were spin degenerate in nature. However on adding more interactions we obtain three poles only. What happened to the other three poles? To understand this, we have made a study of the movement of the poles near 2~GeV region, in the complex plane, as the other interactions are added little by little to the one obtained from the t-channel diagram. In this way we can trace the poles and understand our results better.  
We trace these poles always when the calculations are done without convoluting the loops since otherwise we loose the poles sometimes.

\begin{figure}[h]
\begin{center}
\includegraphics[width=9cm,height=8cm]{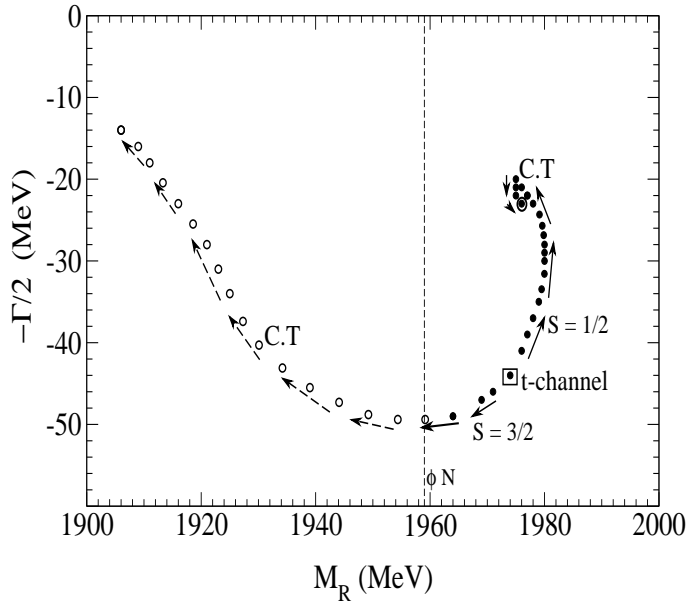}
\caption{   This figure shows how the pole position changes as more diagrams are added to the t-channel vector meson-baryon interaction. The arrows indicate the direction in which the pole moves. The filled (empty) circles represent physical (unphysical) poles.  }
\label{pole1}
\end{center}
\end{figure}

 The spin 1/2, 3/2 pole
at $1974-i44$~MeV, which is found in the calculations carried out by taking the t-channel interaction, is indicated by a boxed point in Fig.~\ref{pole1}.  On adding the contact interaction with very small coupling, to the t-channel,  we find the degenerate poles at $1974-i44$  MeV to split. We find that the spin 1/2 pole moves closer to the real axis and the spin 3/2 pole moves away from the real axis. On  increasing the coupling of the contact interaction from 0 to 1 slowly, we find that the spin 1/2 pole keeps moving towards the real axis whereas the spin 3/2 pole crosses the $\phi N$ threshold and becomes a $\phi N $ virtual state. The direction of the trajectories of the two poles are indicated by arrows in Fig.~\ref{pole1} and a dashed line shows the $\phi N $  threshold. The physical poles are always shown as filled circles and unphysical ones are indicated as empty circles.  The poles obtained by adding the contact interaction with coupling equal to 1, to  the t-channel interaction, are also indicated in Fig.~\ref{pole1} by a label $C.T$.  These poles however move further in the complex plane due to addition of the s- and u-channel diagrams.  As is clear from Fig.~\ref{pole1}, only spin 1/2 pole finally appears as a physical one, the spin 3/2 pole ends up as a $\phi N $ virtual pole.
 
\begin{figure}[h]
\begin{center}
\includegraphics[width=9cm,height=8cm]{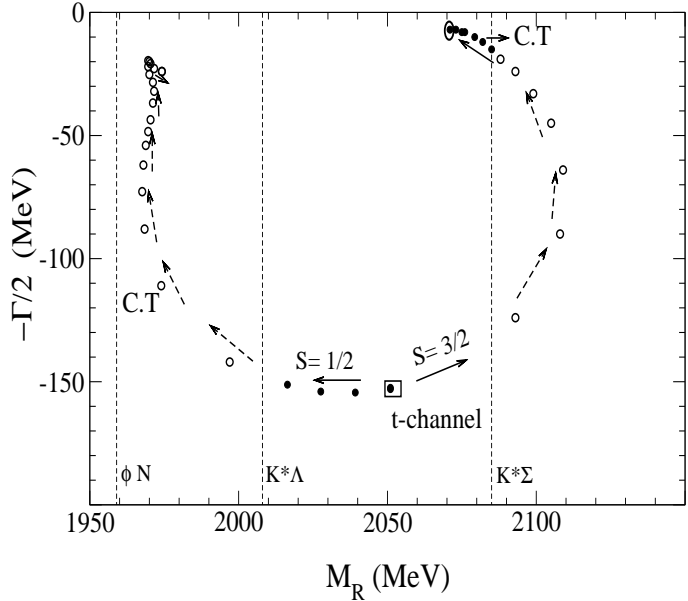}
\caption{   Study of the pole trajectories for yet another pole found in the t-channel calculations. This figure shows the thresholds of $\phi N, K^* \Lambda$ and $K^*\Sigma$ channels as dashed lines.  The meaning of filled and empty circles is same as in Fig.~\ref{pole1}. }
\label{pole2}
\end{center}
\end{figure}

Let  us study yet another pole, the one we find at $2051-i153$~MeV when the calculations are done using the t-channel interaction. This pole is shown as a boxed point in Fig.~\ref{pole2}. The conventions followed in this figure are the same as those in Fig.~\ref{pole1}. Once again we add the contact interaction with a very small coupling to the t-channel and vary this coupling from 0 to 1.  In this case we see that spin 3/2 pole starts moving closer to the real axis, turns into a unphysical pole (appearing on the first Riemann sheet for the $K^* \Sigma$ channel) and later appears as a physical pole below the $K^* \Sigma$ threshold. The spin 1/2 pole also starts moving closer to the real axis but turns into a $K^* \Lambda$ virtual pole. In Fig.~\ref{pole2}, various thresholds have been marked as dashed lines and the corresponding channels are labeled against these lines. Thus in this case too, we find only one pole to end up as a physical pole, the one with spin 3/2.  

This explains why we start with more poles but end with fewer ones. This is so because we find some poles to become unphysical due to addition of more diagrams to the t-channel. However, it should be mentioned that the virtual poles shown in Figs.~\ref{pole1} and \ref{pole2} might become  physical bound states if, for instance, more attraction is added to the system by adding more coupled channels.

\section{Summary}
This work deals with finding of dynamical generation of resonances in the coupled systems made of vector mesons and octet baryons. The leading order
contributions to the scattering equations have been obtained from a sum of diagrams corresponding to a vector meson exchange in the t-channel, octet baryon exchange in s-, u-channels and a contact interaction obtained within the hidden local and chiral symmetries. It is found that the sum of these diagrams give rise to an interaction which has a similar structure as the nucleon-nucleon interaction. However, we find a posteriori  that the  s-channel diagrams are negligibly weak in the present case. The t-channel gives a central potential and the contact interaction and u-channel exchange lead to a spin-spin interactions. A solution of scattering equations results in finding of some resonances which can be identified with known states listed in Ref.~\cite{pdg} as summarized in Table~\ref{summary}. 

\begin{table}[htbp]
\caption{  Summary of the results obtained in the present work.} \label{summary}
\begin{ruledtabular}
\begin{tabular}{cc}
Pole positions  &Corresponding known states   \\
\hline
$1637 - i35$~MeV &  $N^* (1700)~D_{13}$ \\
 $2071- i70$~MeV& $N^* (2080)~D_{13}$ \\
$1977 - i22$~MeV&$N^*(2090)~S_{11}$\\
$2006 -i112$~MeV&$\Delta (1900)~S_{31}$\\
\end{tabular}
\end{ruledtabular}

\end{table}

\section{Acknowledgements}
We are grateful to A.~Mart\'inez~Torres for reproducing all the interactions and for various discussions. This work is partly  supported  by the Grant-in-Aid for Scientific Research on Priority Areas titled ÒElucidation of New Hadrons with
a Variety of Flavors" (E01: 21105006 for K.P.K and A.H) and (22105510 for H.~N) and the authors acknowledge the same. 
\bibliography{basename of .bib file}

\end{document}